\documentclass[aps,draft,twocolumn]{revtex4}

\input epsf

\begin{document}

\title{
Effect of the intrinsic Josephson coupling on the pancake lattice in layered
superconductors.
}
\author{A.V. Rozhkov}

\affiliation{
Institute for Theoretical and Applied Electrodynamics JIHT RAS,
Moscow, ul. Izhorskaya 13/19, 127412, Russian Federation
}

\begin{abstract}
We study the pancake lattice system induced by application of transverse
magnetic filed to the layered superconductor. A simple statistical field
theory for the pancake lattice is derived. It incorporates effects of the
magnetic interaction between pancakes as well as those of the interlayer
Josephson coupling. The proposed description enables us to estimate the
pancake crystal melting temperature from above. Also, it allows us to 
compare directly the effects of the magnetic and Josephson energies on
the statistical properties of the pancake lattice. We demonstrate that even
for such an anisotropic material as BSCCO the pancake interaction induced
by the Josephson coupling is of the same order as the magnetic interaction.
\end{abstract}

\maketitle
\hfill
\section{Introduction}
Magnetic field penetrates a sample of layered superconductor forming
pancake vortices in every layer of the sample. Calculation of statistical
properties of
the pancake matter is a very complicated question for several reasons of
which the following is the most important.
The pancakes interact through two mechanisms, magnetic and Josephson. While
magnetic mechanism is reduced to pairwise interaction of the pancakes,
potential energy due to the interlayer Josephson coupling cannot be brought
down to such simple form. Available methods lead to a non-local description
\cite{hor_gold} or rely on the duality transformation
\cite{rodriguez,rodriguezII} which maps the pancake system onto a system of
quasi-two-dimensional Coulomb-like plasma. Both approaches are technically
complicated and difficult to generalize. 

Alternatively,
many authors neglect Josephson mechanism assuming that for extremely
anisotropic superconductors its influence is small compared to the effects
of the magnetic interaction
\cite{fangohr,blatter,dodgson,dodgsonII,dodgsonIII}.

In this paper we show that under certain circumstances it is possible to
include Josephson coupling into consideration with the help of general
physical devices. At sufficiently high field the Josephson and magnetic
interlayer coupling get smaller than the intralayer interaction. In such a
situation the system is viewed as a collection of weakly coupled 2D
layers. If pancakes in every layer form a structure with short
range crystalline order both magnetic and Josephson mechanisms could be
accounted for. The resultant model is a version of the sine-Gordon field
theory.  It could be analyzed with the help of usual field theoretical tools.

The proposed approach allows one to find an upper bound for the pancake
lattice melting temperature. Another interesting application is that it
becomes possible to compare contributions of the magnetic and Josephson
interactions to the statistical properties of the pancake matter. Thus, we
could determine under what circumstances it is possible to drop the
Josephson coupling from the description and when it cannot be omitted.

The paper is organized as follows. In Sect. II starting from the vortex
Hamiltonian for the anisotropic superconductor we reproduce the
well-known result for the magnetic potential energy of the pancake lattice.
In Sect. III the energy due to the Josephson coupling is derived. In Sect.
IV we compare Josephson and magnetic interaction energies. The purpose of the
latter comparison is to
determine when it is permissible to ignore the Josephson mechanism and when
it is not. In Sect. V we apply the derived Hamiltonian to the problem of
pancake lattice melting. Sect. VI contains conclusions.

\section{Magnetic coupling}
We start with the vortex Hamiltonian $H$
\cite{koshelev,sudbo_brandt}:
\begin{eqnarray}
H= \frac{\Phi_0^2}{8\pi} \int_{\bf q} \left\{ {\cal K}_z({\bf q})
\left| v_z ({\bf q}) \right|^2
+ {\cal K}_\| ({\bf q}) \left| v_\| ({\bf q}) \right|^2 \right\}, \label{H}\\
{\cal K}_z ({\bf q}) = \frac{ 1 + \lambda_c^2 q^2 }{ \left( 1 +
\lambda_{ab}^2 q^2 \right) \left( 1 + \lambda_c^2 q_\|^2 + \lambda_{ab}^2
q_z^2 \right)},\\
{\cal K}_\| ({\bf q}) = \frac{1}{ 1 + \lambda_c^2 q_\|^2 + \lambda_{ab}^2
q_z^2 }.
\end{eqnarray}
The symbol $\int_q \ldots$ stands for $\int (2\pi)^{-3} d^3 q \ldots$;
vector field $\bf v$ denotes vorticity. 
The component $v_z$ in the $n$th superconducting layer coincides with the
pancake density in that layer:
\begin{eqnarray}
v_z ({\bf r}, n) = \rho ({\bf r}, n) = \sum_\mu \delta^{(2)}\left(
{\bf r} - {\bf r}_\mu(n)\right),
\end{eqnarray}
where ${\bf r}_\mu(n)$ is the position of the $\mu$th vortex in the $n$th
layer.

The parallel components of the vorticity ${\bf v}_\|$ correspond to the
Josephson vortices between superconducting layers. The Josephson vorticity
between $n$th and $(n-1)$th layer is equal to:
\begin{eqnarray}
{\bf v}_\| ({\bf r}, n) = \frac{1}{d_c}\sum_\mu 
\{{\bf r}_{\mu}(n) - {\bf r}_{\mu}(n-1)\} \delta_\mu ({\bf r}),
\label{J_vort}\\
\delta_\mu ({\bf r}) = \\
\int_0^1 d\sigma \delta^{(2)} \left( {\bf r} -
{\bf r}_\mu (n-1) - [ {\bf r}_\mu (n) - {\bf r}_\mu (n-1) ]\sigma  \right).
\nonumber
\end{eqnarray}
Here $d_c$ is the spacing between neighboring layers, ${\bf r}$ is 2D
coordinate vector. Function $\delta_\mu$
is defined in such a way that it is localized on a straight line connecting
points ${\bf r}(n)$ and ${\bf r}(n-1)$. Physically, it is non-zero on the
Josephson vortex which connects a pancake in $n$th layer and another
pancake in $(n-1)$th layer.

Since a vortex cannot terminate inside the superconductor we have
\begin{eqnarray}
{\rm div}\, {\bf v} = 0,
\end{eqnarray}
or, equivalently, in Fourier space
\begin{eqnarray}
q_z v_z = - {\bf q}_\| {\bf v}_\|. \label{div}
\end{eqnarray}
Let us rewrite the kernel ${\cal K}_z$ in the following manner:
\begin{eqnarray}
{\cal K}_z ({\bf q}) = {\cal K}_{\rm mag} + \delta{\cal K}_{\rm J},\\
{\cal K}_{\rm mag} = \frac{1}{\lambda_{ab}^2 q_\|^2} \left( 1 -
\frac{1}{1 + \lambda_{ab}^2 q^2} \right),\\
\delta {\cal K}_{\rm J} = - \frac{q_z^2}{q_\|^2 \left( 1 + \lambda_c^2 q_\|^2
+ \lambda_{ab}^2 q_z^2 \right)}.
\end{eqnarray}
Note, that ${\cal K}_{\rm mag}$ contains no $\lambda_c$. In other
words, it does not depend on the Josephson coupling between layers. It
describes magnetic interaction between pancakes. On the contrary, kernel
$\delta {\cal K}_{\rm J}$ vanishes when the Josephson energy vanishes.

Now, using 3D-transversality of the vorticity (eq.(\ref{div})), we establish
the following:
\begin{eqnarray}
\delta {\cal K}_{\rm J} \left| v_z \right|^2 = - \frac{1}{ 1 + \lambda_c^2
q_\|^2 + \lambda_{ab}^2 q_z^2 } \frac{\left| 
\left({\bf q}_\| {\bf v}_\|\right) \right|^2 }{q_\|^2}.
\end{eqnarray}
This allows us to rewrite the Hamiltonian (\ref{H}):
\begin{eqnarray}
H = \frac{\Phi_0^2}{8\pi} \int_q {\cal K}_{\rm mag} \left| v_z \right|^2 +
{\cal K}_{\rm J} \left| v_{\| \rm T} \right|^2,\\
{\cal K}_{\rm J} = \frac{1}{1 + \lambda_c^2 q_\|^2 + \lambda_{ab}^2 q_z^2},
\\
{\bf v}_{\| \rm T} = {\bf v}_\|  - {\bf q}_\|
\frac{({\bf q}_\| {\bf v}_\|)}{q_\|^2}. \label{H2}
\end{eqnarray}
In other words, only the 2D-transverse part ${\bf v}_{\| \rm T}$ of the
Josephson vorticity ${\bf v}_{\|}$ contributes to the interaction energy.

Magnetic part of the energy is the easiest to handle. Imagine that in every
layer the pancakes form 2D crystal with short range hexagonal lattice
characterized by the lattice constant $a_0$.
We assume that it is possible to choose the length scale $\xi_0 \gg a_0$ in
such a way that within $\xi_0$  the lattice translational invariance is
undisturbed. Although, we do not need the exact value of $\xi_0$
for our calculations, yet, our ability to set such a scale is imperative.

If the above prerequisite is met a pancake lattice in $n$th layer could be
described locally by a 2D displacement vector field ${\bf u}_n$. This
field does not change much on the scale $\xi_0$. The pancake density
is expressed as:
\begin{eqnarray}
v_z({\bf r}, n) = \rho_0 \left( 1 + {\rm div}_{2D}\, {\bf u}_n +
\sum_{\bf q_\| \ne 0} e^{i{\bf q_\| r} + i {\bf q_\| u}_n} \right).
\label{vz}
\end{eqnarray}
The sum is upon vectors of the reciprocal lattice. The 2D divergence is
defined as:
\begin{eqnarray}
{\rm div}_{2D} {\bf u} = \frac{\partial u_x}{\partial x} +
\frac{\partial u_y}{\partial y}.
\end{eqnarray}
The quantity $\rho_0$
denotes the average density of the pancakes:
\begin{eqnarray}
\rho_0 = \frac{B}{\Phi_0}.
\end{eqnarray}
Observe, that two first terms of (\ref{vz}) vary slowly on the scale of the
lattice constant $a_0$
while the third term quickly oscillates as function of ${\bf r}$:
\begin{eqnarray}
v_z = \bar \rho + \delta \rho,\\
\bar\rho = \rho_0 \left( 1 + {\rm div}_{2D}\, {\bf u}_n \right),\\
\delta\rho = \rho_0 \sum_{\bf q_\| \ne 0} 
e^{i{\bf q_\| r} + i {\bf q_\| u}_n}.
\end{eqnarray}
We have for the magnetic interaction energy:
\begin{eqnarray}
H_{\rm mag} =
\frac{\Phi_0^2 d_c^2}{8\pi} \sum_{n,m} \int_{ {\bf r},{\bf r'}}
{\cal K}_{\rm mag}({\bf r - r'}, n-m) 
v^{\vphantom{'}}_z v_z' = \\
\frac{\Phi_0^2 d_c^2}{8\pi} \sum_{n,m} \int_{ {\bf r},{\bf r'}}
{\cal K}_{\rm mag}({\bf r - r'}, n-m)\left(\bar\rho
\bar\rho' + \delta\rho \delta\rho'\right), \nonumber 
\end{eqnarray}
where the symbol $\int_{ {\bf r}} \ldots$ stands for
$\int d^2{\bf r}\ldots$.
The term proportional to $\bar\rho \bar\rho'$ describes the statics
of the longitudinal displacements. We will not be interested in them for
they are suppressed \cite{glazman_koshelev}. The remaining parts are:
\begin{eqnarray}
H_{\rm mag} = 
\frac{\Phi_0^2}{8\pi} \rho_0^2 d_c^2 \times \qquad\label{Hmag0}\\
\int_{ {\bf r} {\bf r'}} \left\{ \sum_{n}
{\cal K}_{\rm mag}({\bf r - r'}, 0)
\sum_{\bf q_\| \ne 0}
e^{i{\bf q_\|} ({\bf r - r'}) + i {\bf q_\|}({\bf u}_n^{\vphantom{'}}
- {\bf u}_n') } + \right. \nonumber\\
\left. \sum_{n \ne m}
{\cal K}_{\rm mag}({\bf r - r'}, n-m)
\sum_{\bf q_\| \ne 0}
e^{i{\bf q_\|} ({\bf r - r'}) + i {\bf q_\|}({\bf u}_n^{\vphantom{'}}
- {\bf u}_{m}') } \right\}. \nonumber 
\end{eqnarray}
The notation
\begin{eqnarray}
{\bf u}_n' = {\bf u}_n ({\bf r}')\label{notation}
\end{eqnarray}
was adopted in the formula above. To avoid clutter we omit the subscript
`$T$' which denoted the transverse displacement mode: we write here
${\bf u}_n$ rather than ${\bf u}_{nT}$. We will continue to do so, yet, one
has to remember that the displacement vector field we consider is pure shear.
 
The first term in the above expression gives usual $C_{66}$:
\begin{eqnarray}
C_{66} = \frac{\Phi_0^2 d_c}{(8\pi \lambda_{ab} )^2} \rho_0.
\end{eqnarray}
To proceed
further with the second term we need to calculate ${\cal K}_{\rm mag}$:
\begin{eqnarray}
{\cal K}_{\rm mag} ({\bf q}_\|, n) = \frac{1}{\lambda_{ab}^2 q_\|^2}
\int_{q_z} e^{iq_z d_c n} \left( 1 - \frac{1}{ 1 + \lambda_{ab}^2 q^2 }
\right) = \\
\frac{1}{\lambda_{ab}^2 q_\|^2} \left( d_c^{-1} \delta_{n,0} -
\frac{{\exp}\left\{-|n| d_c {\sqrt{\lambda_{ab}^{-2} + q_\|^2}} \right\}}
{2 \lambda_{ab}^2 \sqrt{  \lambda_{ab}^{-2} + q_\|^2 }} \right).\nonumber
\end{eqnarray}
If $\lambda_{ab} q_\| \gg 1$ which is true for $B>H_{c1}$ the above
expression for ${\cal K}_{\rm mag}$ could be simplified:
\begin{eqnarray}
{\cal K}_{\rm mag} \approx
\frac{1}{\lambda_{ab}^2 q_\|^2} \left( d_c^{-1} \delta_{n,0} -
\frac{e^{-q_\|d_c |n|}}{2\lambda_{ab}^2 q_\| } \right). \label{Kmag}
\end{eqnarray}
Thus, we get for the second term in (\ref{Hmag0}):
\begin{eqnarray}
\sum_{\bf q_\| \ne 0}\int_{ {\bf r}
{\bf r'}} {\cal K}_{\rm mag} ({\bf r - r'}, n-m)
e^{i{\bf q_\|} ({\bf r - r'}) + i {\bf q_\|}({\bf u}_n^{\vphantom{'}}
- {\bf u}_{m}') } \approx \\
\sum_{\bf q_\| \ne 0}\int_{{\bf r}} 
e^{ i {\bf q_\|}({\bf u}_n({\bf r}) - {\bf u}_{m}({\bf r})) } 
\int_{\Delta {\bf r}} {\cal K}_{\rm mag}(\Delta {\bf r}, n-m)
e^{i{\bf q_\|} \Delta{\bf r } } = \nonumber \\
\sum_{\bf q_\| \ne 0}\frac{e^{-q_\| d_c |n-m|}}{2 \lambda_{ab}^4 q_\|^3}
\int_{{\bf r}} 
e^{ i {\bf q_\|}({\bf u}_n - {\bf u}_{m}) }.
\nonumber
\end{eqnarray}
Here we neglected the weak dependence of ${\bf u}({\bf r} + \Delta{\bf r})$
on $\Delta{\bf r}$ since $|\Delta r| \sim a_0 \ll \xi_0$. Finally:
\begin{eqnarray}
H_{\rm mag}&=&\sum_n \int_{{\bf r}} \frac{C_{66}}{2} |\nabla u_n|^2
-\label{Hmag1}\\
&&\sum_{n\ne m} \frac{\Phi_0^2 \rho_0^2 d_c^2}{16\pi\lambda_{ab}^4 q_0^3}
\int_{{\bf r}}  \sum_{{\bf q}_\| \ne 0} 
\frac{e^{-q_\| d_c |n-m|}}{q_\|^3/q_0^3}
e^{ i {\bf q_\|}({\bf u}_n - {\bf u}_{m}) }.
\nonumber
\end{eqnarray}
Here $q_0$ is the magnitude of the smallest reciprocal lattice vectors:
\begin{eqnarray}
q_0 = \frac{4\pi}{\sqrt{3}a_0},\\
q_0^2 = \frac{8\pi^2}{\sqrt{3}} \rho_0.
\end{eqnarray}
Properties of the Hamiltonian similar to (\ref{Hmag1}) was investigated in 
\cite{fangohr,blatter,dodgson,dodgsonII,dodgsonIII}.

At finite temperature the above expression could be simplified. It is
enough to notice that at $T>0$ contribution of a term $\exp(i{\bf q}_\|
({\bf u}_n - {\bf u}_{m}))$ is proportional to $\exp (-q_\|^2 \langle
{\bf u}_n^2 \rangle/2 )$. Thus, when fluctuations of the lattice are
substantial it is permissible to retain in (\ref{Hmag1})
lowest $|{\bf q}_\||$ terms only. Consequently, it is not necessary to sum
over all possible ${\bf q}_\|$ in the above formula. It is sufficient to
keep only six terms corresponding to six elementary reciprocal lattice
vectors whose absolute values are equal to $q_0$. Namely, the last term of
(\ref{Hmag1}) might be written as:
\begin{eqnarray}
&&
\frac{\Phi_0^2 \rho_0^2 d_c^2}{16\pi \lambda_{ab}^4 q_0^3}\times \label{Hmag2}
\\
&&\qquad\sum_{m\ne n}\int_{{\bf r}}  \sum_{|{\bf q}_\| | = q_0} 
{e^{-q_0 d_c |n-m|}}
\cos (  {\bf q_\|}({\bf u}_n - {\bf u}_{m}) ). \nonumber
\end{eqnarray}
Our model becomes a vector version of a sine-Gordon field theory.

\section{Josephson coupling}
Now we add the Josephson coupling to our Hamiltonian. This means that we
have to include the Josephson vorticity as well. At first it seems like an
impossible task since we have to account for the Josephson vorticity
fluctuations. We will demonstrate that under rather broad conditions these
fluctuations could be included by simple renormalization of the Josephson
coupling parameter.
\begin{figure} [h]
\centering
\leavevmode
\epsfysize=6cm
\epsfbox{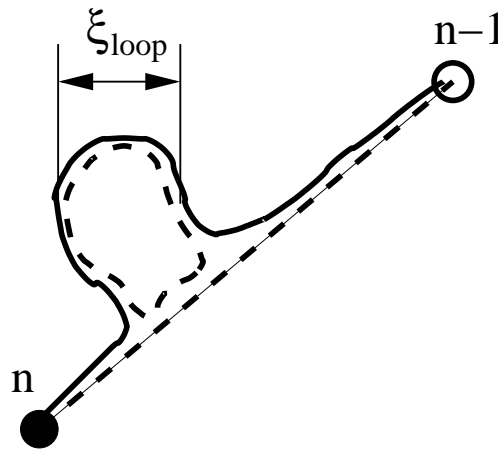}
\caption[]
{\label{fig1} 
A fluctuating Josephson vortex (solid line) connects two pancakes, one in
the $n$th layer (closed circle), another in the $(n-1)$th layer (open
circle). The deformed vortex is decomposed into a straight line and a loop
(both broken lines). The characteristic size of the loop is $\xi_{\rm loop}$.
}
\end{figure}

On fig.1 an example of Josephson vorticity fluctuation is shown. In the
absence of fluctuations two pancake vortices in neighboring layers would be
connected by a straight Josephson vortex. Fluctuations create an outgrowth
on the vortex line. The size of the outgrowth is $\xi_{\rm loop}$.
The deformed vortex is
represented by a solid line. The deformed vortex could be ``decomposed" into
a vortex loop and a straight vortex connecting two pancakes. Thus, the
problem of the Josephson vorticity fluctuations could be reduced to the
problem of the thermally induced vortex loop gas.

Thermal fluctuations infuse superconductor 
with loops of Josephson vortices of typical size $\xi_{\rm loop}$.
 This size $\xi_{\rm loop}$ diverges at
the superconducting transition temperature $T_c$ and vanishes at $T=0$. It is
finite at $T_c > T >0$.
Loosely speaking, $\xi_{\rm loop}$ characterizes a spacial scale above
which ``mean-field" Josephson coupling could be defined and fluctuations
are unimportant. Therefore, if we are in the regime 
\begin{eqnarray}
\xi_{\rm loop} < a_0 \label{fluctI}
\end{eqnarray}
the fluctuations might be omitted. In such a case fluctuations result in
small vibrations of the Josephson vortex. The only consequence
of the fluctuations would be renormalization of effective value of the
Josephson coupling, which is equivalent to renormalization of the
penetration depth from its bare value $\lambda_c ^B$ to experimentally
measurable value of $\lambda_c $.

Of course, it is desirable to generalize the argumentation beyond
(\ref{fluctI}).
How this could be done is discussed in Appendix \ref{energy_of_loop}.

Once we settle the issue of the fluctuations we could proceed with the
derivation of the Josephson contribution to the energy of the pancake
system.

\begin{figure} [h]
\centering
\leavevmode
\epsfysize=6cm
\epsfbox{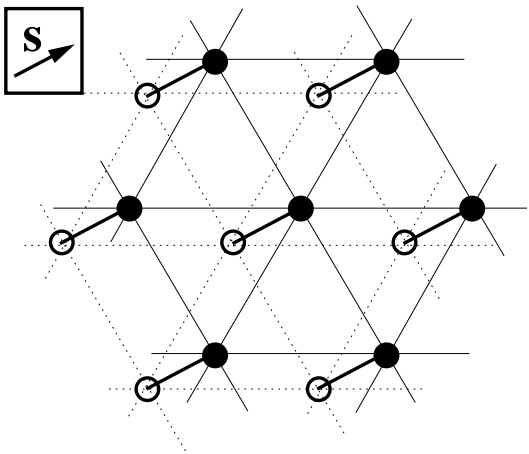}
\caption[]
{\label{fig2} 
The Josephson vorticity generated by the relative displacement of pancake
lattices in the neighboring layers. A hexagonal pancake lattice in the
$n$th layer
(closed circles) is shifted with respect to the lattice in the $(n-1)$th
layer (open circles). Vector ${\bf s}$, eq.(\ref{vect_s}), is shown in a
box.
}
\end{figure}

\begin{figure} [b]
\centering
\leavevmode
\epsfysize=6cm
\epsfbox{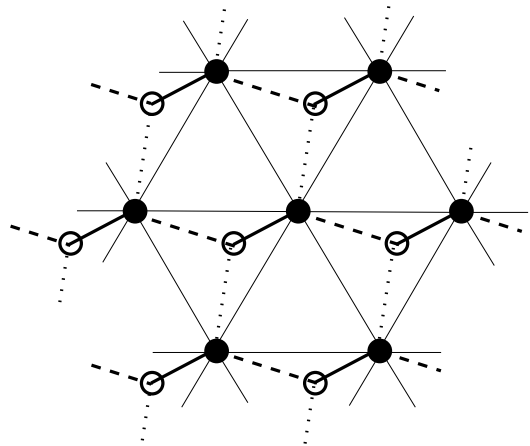}
\caption[]
{\label{fig3} To the definition of vector ${\bf g}$. When two pancake
lattices in two neighboring layers are shifted with respect to each other
pancakes got connected by Josephson vortices in such a way as
to minimize the length of the latter. The optimal choice is shown by solid
lines. Two other possible connections (broken line and dotted line)
require longer vortices which, obviously, costs more energy.
}
\end{figure}

As in the previous Section we assume that in every layer the pancakes form a
2D structure with (at least) short range crystalline order. Remember also
that the longitudinal displacements of the pancakes are suppressed. In
absence of the longitudinal displacements the
Josephson vorticity is a periodic function of vector ${\bf s}_n$ defined as
(fig.\ref{fig2}):
\begin{eqnarray}
{\bf s}_n ({\bf r}) = {\bf u}_n ({\bf r}) - {\bf u}_{n-1} ({\bf r}).
\label{vect_s}
\end{eqnarray}
Two-dimensional vector field ${\bf v}_\|({\bf r}, n)$ could be
expressed as a sum of smooth and oscillating parts: 
\begin{eqnarray}
{\bf v}_\| = \bar{\bf v}_\| + \delta {\bf v}_\| ,
\end{eqnarray}
where the smooth part is equal to:
\begin{eqnarray}
\bar {\bf v}_\| = \frac{\rho_0}{d_c} {\bf g}({\bf s}_n).\label{vort}
\end{eqnarray}
Vector function ${\bf g}$ is defined like so:
\begin{eqnarray}
{\bf g} ({\bf s}) = {\bf s} - {\bf l}({\bf s}).
\end{eqnarray}
The lattice vector ${\bf l}({\bf s})$ is chosen to deliver 
minimum to the expression $\left| {\bf s} - {\bf l} \right|$ (see
fig.(\ref{fig3})). Function ${\bf g}({\bf s})$ is invariant under lattice
translations. For any ${\bf s}$ vector ${\bf g}$ always lies within the
primitive lattice cell. Physically, eq.(\ref{vort}) means that when two
lattices are shifted only slightly with respect to each other $\bar {\bf
v}_\| = \rho_0 {\bf s}/d_c$. However, when ${\bf s}$ does not fit into
the primitive cell the Josephson vortices rearrange themselves to minimize
their own length. In this case $|{\bf g}| < |{\bf s}|$. For example, if ${\bf
s}$ is the lattice vector then ${\bf g} = 0$.

Next, we calculate the oscillating part. Full Josephson vorticity 
field between $n$th and $(n-1)$th layers is given by eq.(\ref{J_vort}) with
${\bf r}_\mu (n-1) = {\bf l}_\mu + {\bf u}_{n-1}$
(${\bf l}_\mu$ is the lattice vector which shows the undisturbed position
of the $\mu$th pancake) 
and ${\bf r}_\mu(n) - {\bf r}_\mu (n-1) = {\bf g}$. The Fourier transform of
the vorticity vector field due to a single vortex $\mu$ is:
\begin{eqnarray}
{\bf v}_{\|\mu}({\bf q}_\|,n) = \frac{{\bf g}({\bf s}_n)}{d_c }
\int_0^1 d\sigma e^{ i{\bf q}_\| \left( {\bf u}_{n-1} + {\bf l}_\mu
+ {\bf g}({\bf s}_n) \sigma\right) } = \nonumber \\
\frac{{\bf g}({\bf s}_n)}{i ({\bf q}_\| {\bf g})d_c} \left(
e^{i{\bf q}_\| ({\bf u}_{n-1} + {\bf g})}
- e^{i {\bf q}_\| {\bf u}_{n-1}  } \right)e^{i{\bf q}_\| {\bf l}_\mu}. 
\end{eqnarray}
The total vorticity is:
\begin{eqnarray}
{\bf v}_\|({\bf q}_\|, n) = \sum_\mu {\bf v}_{\|\mu} ({\bf q}_\|,n).
\end{eqnarray}
It is non-zero only when ${\bf q}_\|$ is the reciprocal lattice vector. If
${\bf q}_\|$ belong to the reciprocal lattice, one prove
\begin{eqnarray}
e^{ i{\bf q}_\| {\bf g}({\bf s}_n)} =
e^{ i{\bf q}_\| {\bf s}_n}.
\end{eqnarray}
Thus, we have for the oscillating part:
\begin{eqnarray}
\delta {\bf v}_\| ({\bf r},n) = \\
\frac{\rho_0}{d_c} \sum_{\bf q_\| \ne 0}
\frac{\bf g(s_{\it n})}{i\bf q_\| g(s_{\it n})} \left(
e^{i{\bf q}_\| {\bf u}_n({\bf r})} -
e^{i {\bf q}_\| {\bf u}_{n-1}({\bf r})} \right)
e^{- i{\bf q}_\| {\bf r}},\nonumber
\end{eqnarray}
where the sum runs over the reciprocal lattice vectors. The above
expression is explicitly periodic with respect to both ${\bf u}_n$ and
${\bf u}_{n-1}$. This means that if one of the layers is shifted by a
lattice vector the Josephson vorticity remains unaffected by such a
shift. 

To calculate the energy we need the transverse component of the Josephson
vorticity:
\begin{eqnarray}
\delta {\bf v}_{\| T} ({\bf r},n)&=&
\frac{\rho_0}{d_c} \sum_{\bf q_\| \ne 0}
\frac{{\bf q}_\|[{\bf q}_\| \times \bf g(s_{\it n})]}
{iq_\|^2(\bf q_\| g(s_{\it n}))}\times\\
&&\left( e^{i{\bf q}_\| {\bf u}_n({\bf r})} -
e^{i {\bf q}_\| {\bf u}_{n-1}({\bf r})} \right)
e^{- i{\bf q}_\| {\bf r}}.\nonumber
\end{eqnarray}
(Note: in 2D vector product is a scalar, not a vector.)
As for $\bar{\bf v}_\|$, it is transverse as long as ${\bf u}_n$'s are
transverse.

We are in position now to write down the Josephson energy:
\begin{eqnarray}
H_{\rm J} = H_{\rm J1} + H_{\rm J2} + H_{\rm J3} ,\\
H_{\rm J1} = \frac{\Phi_0^2 \rho_0^2}{8\pi}
\sum_{n,m} \int_{\bf r, r'}
{\cal K}_{\rm J}({\bf r - r'}, n-m) {\bf g} ({\bf s}_n) 
{\bf g} ({\bf s}'_{m}), \label{HJ1}\\
H_{\rm J2} = \frac{\Phi_0^2 \rho_0^2}{8\pi}
\sum_{n} \int_{\bf r} \sum_{\bf q_\| \ne 0}
{\cal K}_{\rm J} ({\bf q}_\|, 0) 
\left| \frac{[{\bf q}_\| \times \bf g(s_{\it n})]}
{q_\|(\bf q_\| g(s_{\it n}))}\right|^2 \times \label{HJ2}\\
\left( 2 - 2\cos {\bf q}_\|{\bf s}_n\right), \nonumber\\
H_{\rm J3} = \frac{\Phi_0^2 \rho_0^2}{8\pi}
\sum_{n \ne m} \int_{\bf r} \sum_{\bf q_\|\ne 0}
{\cal K}_{\rm J} ({\bf q}_\|, n-m) \times \nonumber \\
\frac{[{\bf q}_\| \times \bf g(s_{\it n})]
[{\bf q}_\| \times \bf g(s_{\it m})]}
{q^2_\|(\bf q_\| g(s_{\it n}))(\bf q_\| g(s_{\it m}))} 
\times \nonumber \\
\left( e^{i{\bf q}_\| {\bf u}_n }-
e^{i {\bf q}_\| {\bf u}_{n-1}} \right)
\left( e^{-i{\bf q}_\| {\bf u}_{m}} -
e^{-i {\bf q}_\| {\bf u}_{m-1}} \right),
\end{eqnarray}
where we used the notation
\begin{eqnarray}
{\bf s}'_{n} = {\bf u}_{n}({\bf r}') - {\bf u}_{n-1}({\bf r}').
\end{eqnarray}
The term $H_{\rm J1}$ describes the kinetic energy of supercurrents induced
by the Josephson vorticity. The kernel
${\cal K}_{\rm J}(\Delta{\bf r}, n)$ is non-zero for $|\Delta {\bf r} |
\ll \lambda_c $ (see Appendix \ref{K}) since such currents spread over large
regions. Therefore, two Josephson vortices apart from each other interact
via these currents. The second term is purely local. It describes the
increase of the free energy of an interlayer Josephson junction 
due to insertion of vortices. The third term describes similar effect: the
modification of the free energy of an interlayer Josephson junction by a
vorticity in another layer. As we will show below, this effect is small.

At low temperature, where it is possible to write ${\bf g}({\bf s}) \approx
{\bf s}$, Hamiltonian $H_{\rm J}$ is responsible for the Josephson
contribution to the tilt modulus $C_{44}$ of the vortex lattice
\cite{glazman_koshelev}. The first term corresponds to the $B^2$ piece of
$C_{44}$ while the second and the third terms contribute to the so-called
single-vortex part ($\propto B$) of the tilt modulus. 

At sufficiently high temperature we cannot approximate ${\bf g}$ by ${\bf
s}$. Instead we must follow different type of analysis.
The term
\begin{eqnarray}
&&\int_{\bf r, r'}
{\cal K}_{\rm J}({\bf r - r'}, n-m) {\bf g} ({\bf s}_n) 
{\bf g} ({\bf s}'_{m})
\end{eqnarray}
may be disregarded right away if $|n-m| \geq 2$
since at low Josephson coupling we could neglect correlations of the
displacement fields in different layers: $\langle{\bf u}_n{\bf u}_{m}
\rangle \approx 0$. Thus, for $|n-m| \geq 2$ 
\begin{eqnarray}
{\bf g} ({\bf s}_{n}) {\bf g} ({\bf s}'_{m}) 
\approx 
{\bf g} ({\bf s}_{n}) 
\langle {\bf g} ({\bf s}'_{m}) \rangle +
{\bf g} ({\bf s}'_{m})
\langle {\bf g} ({\bf s}_{n}) \rangle = 0
\end{eqnarray}
since $\langle {\bf g} ({\bf s}) \rangle = 0$. The latter statement is
obviously true for ${\bf g}$ is an odd function. 

We have:
\begin{eqnarray}
H_{\rm J1} \propto \sum_{m,n} \int_{\bf r, r'}
{\cal K}_{\rm J} {\bf g} ({\bf s}_m^{\vphantom{'}}) 
{\bf g} ({\bf s}'_{n}) = \label{n=m}\\
-\sum_n \sum_{{\bf q}_\|\ne 0 \atop {\bf p}_\| \ne 0} \int_{\bf r r'}
\left\{ {\cal K}_{\rm J}  \hat {\bf g}_{{\bf q}_\|}
\hat {\bf g}_{{\bf p}_\|}
{\rm e}^{i{\bf q}_\|({\bf u}_n - {\bf u}_{n-1}) 
-i{\bf p}_\| ({\bf u}_{n}' - {\bf u}_{n-1}' )}  \right.\nonumber\\
- \left. {\cal K}_{\rm J}  \hat {\bf g}_{{\bf q}_\|}
\hat {\bf g}_{{\bf p}_\|}
{\rm e}^{i{\bf q}_\|({\bf u}_{n+1} - {\bf u}_{n}) 
+i{\bf p}_\| ({\bf u}_{n}' - {\bf u}_{n-1}' )} \right\}, \nonumber
\end{eqnarray}
where notation (\ref{notation}) is used
and $\hat {\bf g}_{{\bf q}_\|} = -\hat {\bf g}_{-{\bf q}_\|}$ are Fourier
transform coefficients of the odd periodic function ${\bf g}({\bf s})$. As
one notices from this formula, $H_{\rm J1}$ is a product of
four exponents of the form $\exp (i{\bf q}_\| {\bf u} )$. It will be
shown that
$H_{\rm J2}$ depends on two exponents only. At high enough temperatures
$H_{\rm J1}$ may be neglected in comparison with $H_{\rm J2}$. This is
because $\langle H_{\rm J2} \rangle$ is proportional to 
$\langle \exp (i{\bf q}_\| {\bf u}_\| ) \rangle^2 $ while 
$\langle H_{\rm J1} \rangle$ is proportional to 
$\langle \exp (i{\bf q}_\| {\bf u}_\| ) \rangle^4 $  and the expectation
value of the exponent vanishes when $T$ grows.

One could adopt a more formal approach. As shown in Appendix
\ref{scaling_dim} at sufficiently large $T$ operator $H_{\rm J1}$ becomes
irrelevant in the renormalization group sense while $H_{\rm J2}$ remains
relevant up until higher temperature. In this temperature interval it is
possible to neglect $H_{\rm J1}$: its contribution to the thermodynamics is
purely perturbative while the contribution of the relevant second term is
singular.

The term $H_{\rm J3}$
could be also disregarded for the following reason. The kernel
\begin{eqnarray}
{\cal K}_{\rm J} ({\bf q}) =
\frac{1}{ 1 + \lambda_c^2 q_\|^2 + \lambda_{ab}^2 q_z^2 } \approx 
\frac{1}{ \lambda_c^2 ( q_\|^2 + \gamma^{-2} q_z^2 ) } ,\\
\gamma = \frac{\lambda_c}{\lambda_{ab}},
\end{eqnarray}
may be simplified provided that
the Josephson length is bigger than the lattice constant:
\begin{eqnarray} 
\lambda_{\rm J} = \gamma d_c \gg a_0. \label{small}
\end{eqnarray}
Namely, we can write:
\begin{eqnarray}
{\cal K}_{\rm J} ({\bf q}) \approx \frac{1}{\lambda_c^2 q_\|^2} \left( 1 - 
\frac{q_z^2}{\gamma^2 q_\|^2} \right).
\end{eqnarray}
The term $q_z^2/\gamma^2 q_\|^2$ is smaller than unity as long as
(\ref{small}) is valid. Indeed, $q_z < \pi/d_c$ and $q_\| > a^{-1}_0$. This
means that ${\cal K}_{\rm J} ({\bf q}_\|, n)$ is smaller than 
${\cal K}_{\rm J} ({\bf q}_\|, 0)$.
Therefore, $H_{\rm J3}$ (corresponds to 
${\cal K}_{\rm J} ({\bf q}_\|, n)$) is much less than
$H_{\rm J2}$ (corresponds to ${\cal K}_{\rm J} ({\bf q}_\|, 0)$).

\begin{figure} [b]
\centering
\leavevmode
\epsfysize=6cm
\epsfbox{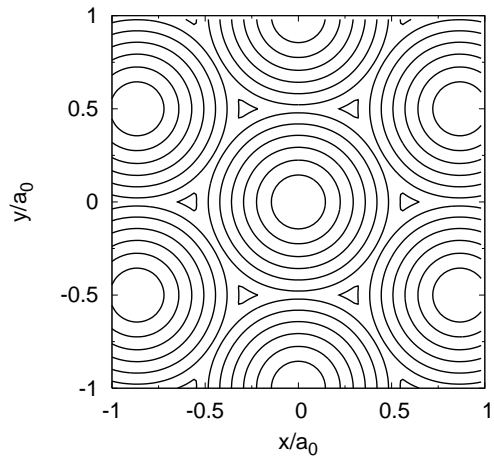}
\caption[]
{\label{fig4} 
Contour plot of the function $h({\bf s}) = h(x,y)$, defined by eq.
(\ref{h_def}). Function $h$ has a minimum at the origin and six maximums at
the corners of the primitive cell.
}
\end{figure}

Thus, we retain $H_{\rm J2}$ only:
\begin{eqnarray}
H_{\rm J} \approx 
\frac{\Phi_0^2 \rho_0^2}{8\pi} \sum_{n} \int_{\bf r}
\sum_{\bf q_\| \ne 0}
{\cal K}_{\rm J} ({\bf q}_\|, 0) 
\left| \frac{[{\bf q}_\| \times \bf g(s_{\it n})]}
{q_\|(\bf q_\| g(s_{\it n}))}\right|^2\times \\
\left( 2 - 2\cos {\bf q}_\|{\bf s}_n\right) ,\nonumber\\
{\cal K}_{\rm J} = \frac{1}{d_c \lambda_c^2 q^2_\|}.
\end{eqnarray}
We rewrite this expression in a more compact way:
\begin{eqnarray}
H_{\rm J} = \frac{3\Phi_0^2 }{256 \pi^5 \lambda_c^2 d_c} \sum_n
\int_{\bf r} h({\bf u}_n ({\bf r}) - {\bf u}_{n-1} ({\bf r})),
\label{HJ_comp}\\
h({\bf s}) =
\sum_{\bf q_\| \ne 0} \frac{q_{0}^4}{q_\|^4}
\left| \frac{[{\bf q}_\| \times \bf g(s)]}
{(\bf q_\| g(s))}\right|^2
\left( 1 - \cos {\bf q}_\|{\bf s}\right) \label{h_def}.
\end{eqnarray}
Function $h$ is dimensionless and periodic with the period of the
hexagonal lattice.
Its contour graph is shown on fig.\ref{fig4}. Due to its periodicity $h$
can be expanded in a Fourier series:
\begin{eqnarray}
h({\bf s}) = \hat h_0 + \sum_{{\bf q}_\| \ne 0} \hat 
h_{\bf q_\|} \cos {\bf q}_\| {\bf s}.
\end{eqnarray}

Using this expansion it is possible to simplify the expression
for $H_{\rm J}$ at finite temperatures with the help of the trick already
discussed at the end of Sec.II:
\begin{eqnarray}
&&H_{\rm J} =\label{HJ}\\
&&\frac{3|\hat h_{q_0}|\Phi_0^2 }{256 \pi^5 \lambda_c^2 d_c} \sum_n
\int_{\bf r} \sum_{|{\bf q}_\| | = q_0}
\{1-\cos ({\bf q}_\|[{\bf u}_n - {\bf u}_{n-1} ])\},\nonumber\\
&&\hat h_{q_0} \approx -3.7,\\
&&\frac{3 |\hat h_{q_0}|}{256 \pi^5} \approx 1.4\times 10^{-4},
\end{eqnarray}
Here the sum over vectors ${\bf q}_\|$ runs only over the shortest of them.
The number
$\hat h_{q_0}$ is the Fourier coefficient of $h({\bf s})$ corresponding to
such reciprocal vectors.
This is the final equation for the Josephson energy of the vortex system. 

\section{Coupling constants}
Let us estimate coupling constants of our Hamiltonian. That way we can
understand what are the largest interaction in our system and 
when the perturbative treatment of interactions is permissible.

First, we calculate
the elastic constant $\varepsilon_{\rm el}$. To find it we measure
area in units of $\rho_0^{-1}$. Thus:
\begin{eqnarray}
\varepsilon_{\rm el} = \frac{C_{66}}{\rho_0} =
\frac{\Phi_0^2 d_c}{64 \pi^2 \lambda_{ab}^2}.
\end{eqnarray}
For BSCCO $\varepsilon_{\rm el} = 182\,~{\rm K}$ assuming:
\begin{eqnarray}
d_c = 1.5\,\text{nm},\\
\lambda_{ab} = 200\, \text{nm}.
\end{eqnarray}
The Josephson constant corresponding to $H_{\rm J2}$:
\begin{eqnarray}
\varepsilon_{\rm J} = \frac{3 \Phi_0^2}{ 256 \pi^5 \lambda_c^2 d_c
\rho_0} |\hat h_{q_0}|.
\end{eqnarray}
The coupling constant corresponding to $H_{\rm J1}$:
\begin{eqnarray}
\varepsilon'_{\rm J} = \frac{\Phi_0^2}{8\pi} \frac{1}{4\pi \lambda_{ab}
\lambda_c } {\bf g}^2_{q_0} \sqrt{\rho_0}.
\end{eqnarray}
Using the estimate for the Fourier coefficients ${\bf g}^2 \approx
1/2q_0^2$ we find:
\begin{eqnarray}
\varepsilon'_{\rm J} \approx \frac{3 \Phi_0^2 a_0}{1024 \pi^4 \lambda_{ab}
\lambda_c }.
\end{eqnarray}
Ratios of these two coupling constants to the elastic energy scale:
\begin{eqnarray}
\frac{\varepsilon_{\rm J}}{\varepsilon_{\rm el}} =
\frac{3 |\hat h_{q_0}|}{ 4\pi^3 \gamma^2 d_c^2 \rho_0} = 
\frac{12 |\hat h_{q_0}|}{ \sqrt{3}\pi^2}\frac{a_0^2}{\lambda_{\rm J} ^2}
\approx 2.6 \frac{a_0^2}{\lambda_{\rm J}^2}
\end{eqnarray}
and
\begin{eqnarray}
\frac{\varepsilon_{\rm J}'}{\varepsilon_{\rm el}} =
\frac{3}{16\pi^2} \frac{a_0}{\lambda_{\rm J} }  \approx 0.02
\frac{a_0}{\lambda_{\rm J}}
\end{eqnarray}
are both smaller than unity when (\ref{small}) holds. This explains the
physical significance of (\ref{small}). The latter is essentially a
criterion for our vortex system to be viewed as a quasi-2D pancake gas. If
(\ref{small}) is violated the system is better described in terms of the
Abrikosov vortices rather than pancakes.

The interlayer magnetic interaction constant could be estimated directly from
(\ref{Hmag1}):
\begin{eqnarray}
\varepsilon_{\rm mag} = 
\frac{\Phi_0^2 \rho_0 d_c^2}{8\pi\lambda_{ab}^4 q_0^3} \frac{1}{q_0 d_c} =
\frac{3\Phi_0^2 d_c}{512 \pi^5 \lambda_{ab}^4 \rho_0}.
\end{eqnarray}
The quantity $1/q_0 d_c$ is an estimate for number of layers coupled by the
magnetic interaction with a given layer.
The ratio of the magnetic and the Josephson constants is:
\begin{eqnarray}
\frac{\varepsilon_{\rm J}}{\varepsilon_{\rm mag}} = \frac{2 |\hat h_{q_0}|
\lambda_{ab}^2}{\gamma^2 d_c^2} =
2 |\hat h_{q_0}| \frac{\lambda_{ab} ^2} {\lambda_{\rm J}^2}
\approx 7.4 \frac{\lambda_{ab} ^2} {\lambda_{\rm J}^2}.
\end{eqnarray}
The last result is important for it allows one to judge when the Josephson
coupling could be neglected and when it must be retained. As one can see
from the above formula, the ratio is independent of the magnetic field and,
in that sense, it is a ``material constant". If we take for BSCCO:
\begin{eqnarray}
\gamma = 300,
\end{eqnarray}
then the ratio equals to 1.5. That
is, even for such an extremely anisotropic material the Josephson coupling
is of the same order as magnetic.

\section{Upper bound for the melting temperature}
\label{upper_bound}
In this section we find the upper bound $T_u$ on a melting temperature
$T_m$. The gist
of the following calculations is that at sufficiently high temperature the
interlayer coupling becomes irrelevant and the system could be thought of
as a collection of 2D layers decoupled from each other. In these layers no
long-range crystalline order is
possible. The temperature found in this fashion is not necessary a true
melting point for melting could occur at even lower temperatures through a
first order phase transition.

In certain situation our upper bound could serve as a reasonable estimate
for the melting temperature. This happens if the Josephson coupling is
sufficiently strong and the difference between the solid and the liquid
phases of the pancake matter is small.  When this difference is exceeded by
the bare Josephson energy scale the phase transition is controlled by the
Josephson energy.  Since the latter vanishes at $T_u$ the melting line lies
close to $T_u$.

We assume that the temperature is large so that Josephson term $H_{\rm J1}$
is irrelevant (see Appendix \ref{scaling_dim}).
Only $H_{\rm J2}$ part of the Josephson coupling and the magnetic coupling,
eq.(\ref{Hmag2}), have to be accounted for. In such a regime it is possible
to get $T_{\rm u}$ using simple perturbative argument which we are ready to
present.

The crystalline phase is characterized by a non-zero order parameter:
\begin{eqnarray}
\Delta = \langle {\rm e}^{i{\bf q}_\| {\bf u}_n} \rangle.
\end{eqnarray}
With the help of $\Delta$ the interlayer interactions (\ref{Hmag2}) and
(\ref{HJ2}) could be written in a mean-field manner as
\begin{eqnarray}
\cos {\bf q_\| (u_{\it n} - u_{{\it m}})} \approx
\Delta \cos {\bf q_\| u_{\it n}}. \label{mf}
\end{eqnarray}
The above transformations reduces our multilayer problem to a problem for a
single layer with the Hamiltonian:
\begin{eqnarray}
H_{\rm mf} = \sum_n \int_{ {\bf r}} \frac{C_{66}}{2} |\nabla u_n|^2 - \alpha
\Delta \sum_{|{\bf q}_\||=q_0}  \cos {\bf q_\| u_{\it n}},\\
\alpha = \rho_0 (\varepsilon_{\rm J} + \varepsilon_{\rm mag}).
\end{eqnarray}
It is presumed that $\alpha$ is small:
\begin{eqnarray}
\alpha \ll C_{66}.
\end{eqnarray}
This guarantee the applicability of approximation (\ref{mf}). It is
equivalent to (\ref{small}).

Let us transform the interlayer interaction term as follows:
\begin{eqnarray}
- \alpha \Delta \cos {\bf q_\| u_{\it n} } \approx
- \alpha \Delta^2
\left( 1 - \frac{1}{2} (\bf q_\| u_{\it n})^2 \right) = \\
\frac{\alpha \Delta^2}{2}
({\bf q_\| u}_n)^2 = \frac{\tilde \alpha}{2} u_n^2,\nonumber\\
\tilde \alpha = \alpha q_0^2 \langle \cos {\bf q_\| u_{\it n} } \rangle^2 
= \alpha q_0^2 \exp \left( - \langle
({\bf q_\| u}_n)^2 \rangle \right).
\end{eqnarray}
After such substitution the effective Hamiltonian becomes quadratic. Thus,
the problem could be solved exactly. However, when solving it we must keep in
mind the self-consistency condition:
\begin{eqnarray}
\langle ({\bf q_\| u}_{n} )^2 \rangle = \int_{\bf k} \frac{T ({\bf q_\|
k})^2}{k^2(C_{66} k^2 + \tilde \alpha)} =\label{self-cons} \\
\frac{q_0^2 T}{8\pi C_{66}} \left[ \ln \frac{C_{66} q_0^2}{\tilde \alpha}
+ {\rm const.} \right]=\nonumber\\
\frac{q_0^2 T}{8\pi C_{66}} \left[ \ln \frac{C_{66} }{\alpha} +
\langle ({\bf q_\| u}_{n} )^2 \rangle  + {\rm const.}\right].\nonumber
\end{eqnarray}
The last equation has positive solution for 
$\langle ({\bf q_\| u}_{n} )^2 \rangle $ only when
\begin{eqnarray}
T < T_u = \frac{8\pi C_{66}}{q_0^2} = \frac{\sqrt{3} \Phi_0^2 d_c}{64 \pi^3
\lambda_{ab}^2}.\label{Tu}
\end{eqnarray}
Above $T_u$ the order parameter $\Delta$ is zero.

Since $\lambda_{ab} $ is a function of temperature itself, to find
$T_u$ it is in fact necessary to solve the equation:
\begin{eqnarray}
T_u =  \frac{\sqrt{3} \Phi_0^2 d_c}{64 \pi^3 \lambda_{ab}^2(T_u)}.
\label{Tu_eq}
\end{eqnarray}
Assume that the temperature dependence of $\lambda_{ab} $ is given by a
phenomenological formula:
\begin{eqnarray}
\lambda_{ab} ^{-2}(T) = \lambda_{ab} ^{-2} (0) ( 1 - T^2/T^2_c ),
\end{eqnarray}
where the superconducting transition temperature $T_c$ equals to 100\,K for
BSCCO. Eq.(\ref{Tu_eq}), thus, gives:
\begin{eqnarray}
T_u \approx 62\,{\rm K}. \label{Tu_BSCCO}
\end{eqnarray}
Let us reiterate, that this result is ``high-field''. Namely, it is derived
under the assumption that (\ref{small}) is satisfied. The latter inequality
is equivalent to:
\begin{eqnarray}
B \gg \frac{\Phi_0}{\lambda_{\rm J}^2}.\label{high_B}
\end{eqnarray}
Thus, there is no contradiction if at low fields
($B < \Phi_0/\lambda_{\rm J}^2$) the melting temperature is higher
than $T_u$ found above.

We can also find the contribution of the interlayer (Josephson and magnetic)
coupling to the total free energy density of the pancake lattice:
\begin{eqnarray}
F_{\rm inter}/S = - \alpha \langle \cos ({\bf q}_\| {\bf u}) \rangle^2 =\\
-\alpha \exp \left( - \langle ({\bf q}_\| {\bf u})^2 \rangle \right) =
-C_{66} \left( \frac{\alpha}{C_{66}} \right)^
\frac{8\pi C_{66}}{8\pi C_{66} - q_0^2 T},\nonumber
\end{eqnarray}
where the following expression for
$\langle ({\bf q}_\| {\bf u})^2\rangle$ was used:
\begin{eqnarray}
\langle ({\bf q}_\| {\bf u})^2 \rangle = 
\frac{q_0^2 T}{8\pi C_{66} - q_0^2 T}
\ln\left(\frac{C_{66}}{\alpha}\right).
\end{eqnarray}
The latter expression is a trivial consequence of (\ref{self-cons}).

Note that $\alpha/C_{66} \sim (\varepsilon_{\rm J} + \varepsilon_{\rm mag})
/\varepsilon_{\rm el}$ and
$q_0^2T/8\pi C_{66} = T/T_u$. Thus, the formula for $F_{\rm inter}$ could be
rewritten:
\begin{eqnarray}
F_{\rm inter}/S \sim - C_{66} \left(\frac{3|\hat h_{q_0}|}{4\pi^3 \gamma^2
d_c^2 \rho_0} + \frac{3}{8\pi^3 \lambda_{ab} ^2 \rho_0} \right)^{1/(1-T/T_u)}.
\end{eqnarray}
Interlayer contribution to the free energy, as expected, vanishes at
$T=T_u$. This result is useful only if the lattice remains stable close to
$T_u$, that is, if $T_m$ is close to $T_u$.

\section{Discussion}
In this paper we proposed a description of the quasi-2D pancake lattice.
The description is a version of an elasticity theory: the Hamiltonian is a
functional of the displacement field only. Such a transparent structure
allows easy generalization. The most obvious extension is for the case of
non-zero pinning. This is an important direction for future studies.

Another advantage of the method is that at hight magnetic field
(eq.(\ref{high_B})) all non-harmonic terms are small. Therefore, consistent
perturbative analytical tools are applicable. 

At present we do not include dislocations and vacancies into our model.
Yet, to certain extent the effect of both dislocations and vacancies may be
accounted for even at this stage.
Unlike the interlayer coupling, these two disorder the lattice. The
interlayer coupling tries to eliminate them by binding into a topologically
neutral pairs. In turn, dislocations and vacancies attempt to destroy the
interlayer
coupling by breaking translational order. Which side wins this competition
could be determined through comparison of corresponding singular corrections
to the free energy. At least conceptually, this is a straightforward task.

The main goal of this paper is the derivation of the model. However, we'd
like to discuss the most elementary of its consequences here. 

First of all, our model allows to compare quantitatively the magnitude of
the magnetic and
Josephson pancake interactions. We have seen above that even for BSCCO
($\gamma \sim 300$) both of these are of the same order. Therefore, models
which neglect the Josephson interaction are only qualitatively
accurate, at best.

Next, let us briefly discuss the application of formula (\ref{Tu_eq}) for
pancake lattice melting in real layered superconductors. Probably, it has
no physical significance for BSCCO due to its extremely small Josephson
coupling. The total interlayer coupling is also small since the magnetic
interaction is of the same order as the Josephson. The transition into the
liquid phase occurs because the free energy of the liquid becomes less
than the free energy of the solid. Small contribution of the interlayer 
interaction could change the melting line slightly but otherwise is
insignificant.

We would like to conjecture cautiously that (\ref{Tu_eq}) may be of
relevance for Pb-doped BSCCO. This superconductor has much smaller
anisotropy $\gamma \sim 70$ \cite{bazilevich}. Presumably, this indicates
that its Josephson coupling parameter is higher than that of pure BSCCO.

In \cite{uspenskaya} Pb-doped BSCCO was studied by means of magneto-optic
technique. A
depinning transition was observed at $T^* \approx 54~{\,\rm K}$.
The authors of \cite{uspenskaya} interpreted the transition as a
pancake lattice melting. Remarkably, they see that this transition is field
independent. This is
a strong argument in favor of (\ref{Tu_eq}) being applicable since $T_u$
does not depend on the magnetic field. Another field-independent melting
transition, 2D dislocation unbinding, is an unlikely candidate for it
occurs at much smaller temperatures. 

Pb-doped BSCCO has $T_c \approx 91$~K and $d_c \approx 1.9\,$~nm. In
\cite{lu} $\lambda_{ab}$ was reported to be of the order of 180~nm. Thus:
\begin{eqnarray}
T_u \approx 68\/{\rm K}. \label{Tu_Pb}
\end{eqnarray}
This is close to the transition at $T^* \approx 54\/$~K reported in
\cite{uspenskaya}. The fact, that
$T_u$ and $T^*$ differs by about 25\% is not by itself discouraging since
penetration depth $\lambda_{ab} $ shows strong sample-to-sample variations
\cite{uspenskaya_private}. It is very well possible that they are
responsible for the deviation of the experimental transition temperature
from its theoretical estimate. Yet, the final judgment about the
applicability of the presented theory is to be postponed until further
investigation.

In conclusion, we proposed a simple model for the pancake lattice. It could
be used to study pinning and thermodynamics of the pancake matter at
sufficiently high fields and temperatures.

\section{Acknowledgments}

Numerous fruitful discussions with A.L. Rakhmanov are appreciated.
The author is grateful for support provided by the Dynasty Foundation,
by RFBR through grant 03-02-16626 and by Russian federal
program "Leading scientific schools" NSh-1694.2003.2.

\appendix

\section{Thermally induced Josephson vorticity}
\label{energy_of_loop}

In this Appendix we will try to estimate the characteristic size
$\xi_{\rm loop}(T)$ of the thermally induced Josephson vortex loops at given
temperature $T$.We will also  discuss what happens when $\xi_{\rm loop}$
grows bigger than $a_0$.

To achieve the first goal we calculate the energy of a single
Josephson vortex loop $E_{\rm loop} (R)$ as a function of its radius $R$ 
and compare this energy to the temperature. We will assume that $\xi_{\rm
loop}$ is determined by the condition
\begin{eqnarray}
E_{\rm loop}(\xi_{\rm loop}) = T. \label{xi_eq}
\end{eqnarray} 
To proceed with the calculation we first find the Fourier transformation of
the vorticity field of a single loop whose radius is $R$:
\begin{eqnarray}
{\bf v}_\|({\bf q}) = \oint_{r=R} e^{i {\bf q}_\| {\bf r}} [\hat {\bf z}
\times d{\bf r} ] = 2\pi R J_1 (q_\| R) \frac{[\hat {\bf z} \times
{\bf q}_\|]} {q_\|}.
\end{eqnarray}
The energy is given by:
\begin{eqnarray} 
E_{\rm loop} (R) =
\frac{\Phi_0^2}{8\pi} R^2 \int_{{\bf q}_\|,q_z} \frac{4 \pi^2
J_1^2(q_\| R) }{ 1 + \lambda_c^2 q_\|^2 + \lambda_{ab}^2 q_z^2 }.
\end{eqnarray}
The integral in this formula could be transformed as follows:
\begin{eqnarray}
I=
\int_{{\bf q}_\|,q_z} \frac{4 \pi^2
J_1^2(q_\| R) }{ 1 + \lambda_c^2 q_\|^2 + \lambda_{ab}^2 q_z^2 } =\\
2 \int_0^{1/\xi} \frac{J_1^2(q_\| R)}{\lambda_{ab} \sqrt{1 + \lambda_c^2
q_\|^2}} {\rm arctg} \left( \frac{\pi \lambda_{ab} }{d_c \sqrt{1 + \lambda_c^2
q_\|^2}} \right) q_\| d q_\|, \nonumber
\end{eqnarray}
where $\xi$ is the superconducting coherence length.
After a substitution $x=q_\| R$ the integral becomes:
\begin{eqnarray}
I= \frac{2}{\lambda_{ab} \lambda_c R}\int_0^{R/\xi} J_1^2(x) {\rm arctg}
\left(\frac{\pi R}{\lambda_{\rm J} x} \right) dx.
\end{eqnarray}
In deriving the above equation we used the approximation $\sqrt{1 +
\lambda_c q_\|^2} \approx \lambda_c q_\|$. This is accurate for $q_\| \gg
\lambda_c^{-1}$ or, equivalently, $R \ll \lambda_c $.

Finally, confining ourselves to the range $\xi \ll R \ll \lambda_{\rm J}$ we
write:
\begin{eqnarray}
I \approx \frac{2}{\lambda_{ab} \lambda_c R} \left(\frac{\pi R}
{\lambda_{\rm J}} \right) \int_0^\infty J_1^2(x) \frac{dx}{x} =
\frac{\pi}{\lambda_c^2 d_c}.
\end{eqnarray}
Therefore:
\begin{eqnarray}
E_{\rm loop} \approx \frac{\Phi_0^2 d_c}{8 \lambda_{ab}^2} \left(
\frac{R}{\lambda_{\rm J}} \right)^2 = \frac{8\pi^3}{\sqrt{3}} T_u 
\left(\frac{R}{\lambda_{\rm J}}\right)^2,
\end{eqnarray}
where $T_u$ is given by (\ref{Tu_eq}).
Using (\ref{xi_eq}) one determines:
\begin{eqnarray}
\xi_{\rm loop} \sim 0.08 \lambda_{\rm J} \sqrt{\frac{T}{T_u}}.
\end{eqnarray}
At $T=T_u$ which is the highest temperature at which the model proposed in
this paper is still
applicable we have 
\begin{eqnarray}
\xi_{\rm loop} \alt 0.08 \lambda_{\rm J}.
\end{eqnarray}
This give us the upper bound for the characteristic size of the
thermally induced Josephson vorticity loops.

As it was discussed in the body of the paper we would like to have
$\xi_{\rm loop} \ll a_0$ (eq.(\ref{fluctI})). Consequently,
\begin{eqnarray}
0.08 \lambda_{\rm J} \ll a_0.
\end{eqnarray}
It is not difficult to notice
that the latter condition is not always compatible with the requirement
$a_0 \ll \lambda_{\rm J}$ which has to be imposed to observe the quasi-2D
effects of the pancake physics (eq.(\ref{small})). What does this mean for
the analysis undertaken in this paper?

The answer to this question is that it is possible to recover virtually all
of the physics described in the paper even if (\ref{fluctI}) is not
satisfied as long as $\xi_{\rm loop} \ll \xi_0$. (The quantity $\xi_0$ was
defined in the paper as a size within which the 2D pancake lattice could be
viewed as almost ideal.) Let us examine how this is done.

Presumably, the violation of (\ref{fluctI}) might lead to problems with
(\ref{HJ_comp}). It was calculated under assumption that it is possible to
define a single-vortex line tension on a scale $\sim a_0$. When vorticity
fluctuation scale $\xi_{\rm loop}$ exceeds $a_0$ it becomes difficult to
assign a well-defined line tension to a Josephson vortex. Thus, the
analysis at the scale of $a_0$ fails. 

It is, however, possible to study the system at the scale of $\xi_0$. At
this scale we need not to be concerned with the properties of a single
Josephson vortex. Instead, we will think in terms of average 
local vorticity which
pierces interlayer Josephson junction.

Imagine first the situation when pancake lattices in the $n$th and
$(n-1)$th layers coincide at ${\bf r} = {\bf r}_{0}$. This implies that
for any ${\bf r}$, $|{\bf r} - {\bf r}_{0}| < \xi_0$, the average Josephson
vorticity between these two layers is absent. Thus, the local density of
the Josephson energy is zero. Now we shift the lattices with respect to
each other: ${\bf s}_n = {\bf u}_n({\bf r}) - {\bf u}_{n-1}({\bf r}) \ne 0$.
The average vorticity $\langle {\bf v}_\|({\bf s})\rangle$ is no longer zero
as well. We have to disentangle its contribution from contribution of the
fluctuation-induced vorticity $\delta {\bf v}_\|$ in the situation where
fluctuations locally are strong:
$\langle |\delta {\bf v}_\|| \rangle > |\langle {\bf v}_\| \rangle |$. To
do so we coarse-grain the system description up to the scale $\xi_{\rm
loop}$. After this coarse-graining $\delta{\bf v}_\| \approx 0$ and the
bare value of the penetration depth
$\lambda_c ^B$ has to be replaced by $\lambda_c $. (The latter is the
penetration depth measured experimentally.) Thus, at the scale
$\xi_0 \gg \xi_{\rm loop}$ the vorticity fluctuations could be neglected 
and it is possible to work with the effective Josephson coupling instead of
including fluctuations explicitly.

The ``mean-field'' vorticity $\langle {\bf v}_\| ({\bf r}) \rangle$ is a
periodic function of ${\bf s}_n$. This implies that the pancake lattice
energy density due
to Josephson coupling is a periodic function of ${\bf s}_n$ as well:
\begin{eqnarray}
H_{\rm J} = \sum_n
\int_{{\bf r}} \varepsilon_{\rm J} \rho_0
\sum_{|{\bf q}_\|| = q_0} 
[ 1 - \cos({\bf q}_\| {\bf s}_n) ]. \label{HJ_est}
\end{eqnarray}
This expression is completely general: it is the lowest ${\bf q}_\|$ terms
of periodic function Fourier series.

The final question now is the estimation of $\varepsilon_{\rm J}$. The
latter, however, is fairly trivial. As it was discussed in the body of the
paper $H_{\rm J2}$ comes from the increase of the interlayer Josephson
junction energy due to non-zero $\langle {\bf v}_\| ({\bf r}) \rangle$.
Thus, we have match (\ref{HJ_est}) and the interlayer Josephson junction
free energy:
\begin{eqnarray}
{\cal F}_{\rm J}({\bf r}_{0}) = \frac{\Phi_0^2}{16 \pi^3 \lambda_c^2 d_c}
[1 - \cos \{\phi_n ({\bf r}_{0})-\phi_{n-1} ({\bf r}_{0})\}]
\end{eqnarray}
When ${\bf s}_n = 0$ both are zero. When $|{\bf s}_n| \sim a_0/2$ the
Josephson junction energy ${\cal F}_{\rm J}$ is
no longer zero but rather equals to 
$\Phi_0^2/16 \pi^3 \lambda_c^2 d_c$. It is because the cos term in the
above expression effectively vanishes due to fast oscillation of the phase
difference. This implies that
\begin{eqnarray}
\varepsilon_{\rm J} \rho_0 \max_{\bf s}\left\{
\sum_{|{\bf q}_\|| = q_0} [ 1 - \cos({\bf q}_\| {\bf s})]\right\}
 \sim \frac{\Phi_0^2}{16 \pi^3 \lambda_c^2 d_c}. \label{eJ_est0}
\end{eqnarray} 
Since maximum of the sum in this formula is equal to 9 the coupling
constant $\varepsilon_{\rm J}$ is:
\begin{eqnarray}
\varepsilon_{\rm J}\rho_0 \sim \frac{\Phi_0^2}{144 \pi^3 \lambda_c^2 d_c} =
2.2\times 10^{-4}  \frac{\Phi_0^2}{ \lambda_c^2 d_c}. \label{eJ_estI}
\end{eqnarray}
Equations (\ref{HJ_est}) and (\ref{eJ_estI}) reproduce (\ref{HJ2}) up to a
numerical constant of order unity.

The derivation of (\ref{HJ2}) developed in this Appendix is more general
than that found in the body of the paper for it does not rely on $\xi_{\rm
loop} \ll a_0$ condition. The advantage of the latter,
however, it its ability to give the numerical constant accurately. The fact
that numerical coefficients of (\ref{HJ2}) and (\ref{eJ_estI}) are very
close is pure luck. Obviously, the estimates like (\ref{eJ_est0}) 
have ``order of magnitude'' precision only. Numerical constant in
(\ref{eJ_estI}) should not be taken too seriously.

We re-derived $H_{J2}$ above. Operator $H_{J1}$ may be re-derived in the
similar fashion.

\section{Scaling dimensions}
\label{scaling_dim}
In this Appendix we determine the scaling dimension (SD) of the operators
(\ref{n=m}) and (\ref{HJ}). We start by solving an auxiliary problem: we
calculate SD of an exponent $\exp(i{\bf q}_\| {\bf u} )$. The easiest way
to find SD is to evaluate the correlation function:
\begin{eqnarray}
\langle e^{i{\bf q}_\| {\bf u}({\bf r})}e^{-i{\bf q}_\| {\bf u}({\bf r}')}
\rangle = e^{-\frac{1}{2} \langle ({\bf q}_\| [{\bf u} - {\bf u}'])^2
\rangle}.
\end{eqnarray}
The correlation function of the displacement field is found:
\begin{eqnarray}
\langle ({\bf q}_\| [{\bf u} - {\bf u}'])^2\rangle =
\int_{{\bf k}} \frac{T({\bf q}_\| {\bf k})^2}{C_{66}k^4} \left( 1 - e^{i{\bf
k} ({\bf r}^{\vphantom{'}} - {\bf r}')} \right).
\end{eqnarray}
The integral over ${\bf k}$ was evaluated in \cite{chaikin_lubensky}
(eq. (9.3.33) and problem (9.9) of the latter reference): 
\begin{eqnarray}
\int_{{\bf k}} \frac{k_{i} k_{j}}{k^4} \left( 1 - e^{i{\bf
k} {\bf r}} \right) = \frac{1}{4\pi}
\delta_{ij} \ln \left(|{\bf r}|/a_0 \right) + \ldots,
\end{eqnarray}
where ellipsis stand for terms which remain bound at $|{\bf r}|
\rightarrow \infty$. Consequently:
\begin{eqnarray}
\langle ({\bf q}_\| [{\bf u} - {\bf u}'])^2\rangle = \frac{Tq_0^2}
{4\pi C_{66}} \ln |{\bf r} - {\bf r}'|/a_0 + \ldots.
\end{eqnarray}
Therefore:
\begin{eqnarray}
\langle e^{i{\bf q}_\| {\bf u}({\bf r})}e^{-i{\bf q}_\| {\bf u}({\bf r}')}
\rangle \propto \left(\frac{a_0}{|{\bf r} - {\bf r}'_\||} \right)^{2T/T_u}.
\end{eqnarray}
From this equation we can read off SD of the exponent:
\begin{eqnarray}
\left[ e^{i{\bf q}_\| {\bf u} } \right] = T/T_u,
\end{eqnarray}
where $T_u$ is given by (\ref{Tu}).
Thus, SD of (\ref{HJ}) is equal to $2T/T_u$ since it is a product of two
exponents. This SD becomes equal to 2 when $T=T_u$. At $T>T_u$ operator
(\ref{HJ}) is irrelevant, otherwise it is relevant. As usual, this mark
the transition point from ordered to disordered phase driven by
(\ref{HJ}).
Eq.(\ref{Tu}) establishes the same fact derived by other means.

As for (\ref{n=m}), its operator density is proportional to:
\begin{eqnarray}
\int \frac{d^2 {\bf R}_\|}{ R_\|} {\bf g}({\bf s}({\bf r})) 
{\bf g}({\bf s}({\bf r} + {\bf R}_\|)).
\end{eqnarray}
Here the factor $1/R_\|$ comes from the kernel ${\cal K}_{\rm J}$. The
kernel ${\cal K}_{\rm J} $ is calculated in Appendix \ref{K}.

SD of ${\bf g}$ is $2T/T_u$ as it is obvious from (\ref{n=m}), SD of the
integration measure is -2, SD of $1/R_\|$ is 1. Totally, we have
\begin{eqnarray}
\left[ \int
\frac{d^2 {\bf R}_\|}{ R_\|} {\bf g}{\bf g} \right] = 4T/T_u - 1.
\end{eqnarray}
Equating this SD with 2 we discover that for $T > 3T_u/4$ operator
(\ref{n=m}) is irrelevant. Therefore, is is permissible to neglect it for
these temperatures. At the same time, operator (\ref{HJ}) remains relevant
up until the temperature has grown to $T_u$. Thus, the discussion of the
Sect.\ref{upper_bound} is valid for
\begin{eqnarray}
\frac{3T_u}{4} < T < T_u.
\end{eqnarray}

\section{Kernel ${\cal K}_{\rm J}$}
\label{K}
In this Appendix we calculate the kernel ${\cal K}_{\rm J} ({\bf r}, 0)$
required for eq.(\ref{n=m}).

We must evaluate the integral:
\begin{eqnarray}
{\cal K}_{\rm J} ({\bf r},0)
= \int_{{\bf k}_\|} \int_{-\pi/d_c}^{\pi/d_c}
\frac{dk_z}{2\pi} \frac{e^{i{\bf k}_\| {\bf r}}}{ 1 + \lambda_c^2 k_\|^2 +
\lambda_{ab}^2 k_z^2 } = \\
= \frac{1}{\pi \lambda_{ab} } \int_{{\bf k}_\|}
\frac{e^{i{\bf k}_\| {\bf r}}}
{\sqrt{ 1 + \lambda_c^2 k_\|^2 }}
{\rm arctg}\left( \frac{\pi \lambda_{ab} }
{d_c \sqrt{ 1 + \lambda_c^2 k_\|^2 }} \right). \nonumber
\end{eqnarray}
In most layered superconductors the screening length $\lambda_c$ is very
big. Therefore, we are interested in distances $r \ll \lambda_c $. For such
$r$ it is possible to neglect unity under the square root and write:
\begin{eqnarray}
{\cal K}_{\rm J} \approx 
 \frac{1}{\pi \lambda_{ab} \lambda_c } \int_{{\bf k}_\|}
\frac{e^{i{\bf k}_\| {\bf r}}}{k_\|}
{\rm arctg}\left( \frac{\pi}
{\lambda_{\rm J} k_\|} \right). 
\end{eqnarray}
Next we perform the angular integration:
\begin{eqnarray}
{\cal K}_{\rm J} \approx 
\frac{1}{2\pi^2 \lambda_{ab} \lambda_c } \int_0^{1/a_0} J_0(k_\| r) 
{\rm arctg}\left( \frac{\pi} {\lambda_{\rm J} k_\|} \right) dk_\| =\\
\frac{1}{2\pi^2 \lambda_{ab} \lambda_c r} \int_0^{r/a_0} J_0(x) 
{\rm arctg}\left( \frac{\pi r} {\lambda_{\rm J} x} \right) dx.\nonumber
\end{eqnarray}
The upper limit in this integral is assumed to be large: $r \gg a_0$.

If $r\ll \lambda_{\rm J}$ then arctg$(\pi r/\lambda_{\rm J} x)$ could be
approximated by $\pi r/\lambda_{\rm J} x$ for almost whole integration
interval. The contribution of the interval where this approximation is
invalid could be evaluated separately:
\begin{eqnarray}
{\cal K}_{\rm J} \approx 
\frac{1}{2\pi^2 \lambda_{ab} \lambda_c r} \left\{ \int_0^
{\pi r/\lambda_{\rm J}} J_0 (x)
{\rm arctg}\left( \frac{\pi r} {\lambda_{\rm J} x} \right) dx\right.\\
\left.+ \int_{\pi r/\lambda_{\rm J}} ^{r/a_0} J_0(x) \frac{\pi r}
{\lambda_{\rm J} x} dx \right\}.
\nonumber
\end{eqnarray}
The first integral is ${\cal O}({\pi r/\lambda_{\rm J}})$. We neglect it.
The second integral must be integrated by parts:
\begin{eqnarray}
\int_{\pi r/\lambda_{\rm J}} ^{r/a_0} \frac{J_0(x)}{x} dx \approx
-\ln \frac{\pi r}{\lambda_{\rm J}} + \int_0^{\infty} J_1(x) \ln x dx.
\end{eqnarray}
Consequently:
\begin{eqnarray}
{\cal K}_{\rm J} \approx \frac{1}{2\pi \lambda_c^2 d_c} \left[ \ln
\frac{\pi r}{\lambda_{\rm J}} + {\rm const.} \right] \text{\ for\ }
r \ll \lambda_{\rm J}.
\end{eqnarray}
In the opposite limit $r \gg \lambda_{\rm J}$ we define a function:
\begin{eqnarray}
f(x) = \int_0^x dy J_0(y),
\end{eqnarray}
and write:
\begin{eqnarray}
{\cal K}_{\rm J} \approx 
\frac{1}{2\pi^2 \lambda_{ab} \lambda_c r} \int_0^{r/a_0} {\rm arctg}
\left(\frac{\pi r}{\lambda_{\rm J} x} \right) d f = \\
\frac{1}{2\pi^2 \lambda_{ab} \lambda_c r} \left( \frac{\pi r}{\lambda_{\rm
J} } \right) \int_0^{r/a_0} \frac{f(x) dx}
{ x^2 + (\pi r/ \lambda_{\rm J} )^2 }.
\end{eqnarray}
The integral may be transformed in the following manner:
\begin{eqnarray}
\int_0^{r/a_0} \frac{f(x) dx} { x^2 + (\pi r/ \lambda_{\rm J} )^2 } =
\int_0^{1} \frac{f(x) dx} { x^2 + (\pi r/ \lambda_{\rm J} )^2 }+\\
\int_1^{r/a_0} \frac{f(x) dx} { x^2 + (\pi r/ \lambda_{\rm J} )^2 }
\nonumber
\approx\\
\int_0^{1} \frac{f(x) dx} { x^2 + (\pi r/ \lambda_{\rm J} )^2 }+
f(\infty) \int_1^{r/a_0} \frac{dx} { x^2 + (\pi r/ \lambda_{\rm J} )^2 }.
\nonumber
\end{eqnarray}
In the last line the first integral is smaller than the second one. Indeed,
the first one is ${\cal O}((\lambda_{\rm J} /r)^2)$ while the second is
${\cal O}(\lambda_{\rm J} /r)$. Therefore:
\begin{eqnarray}
\int_0^{r/a_0} \frac{f(x) dx} { x^2 + (\pi r/ \lambda_{\rm J} )^2 }
\approx\\
f(\infty) \int_1^{r/a_0} \frac{dx} { x^2 + (\pi r/ \lambda_{\rm J} )^2 }
\approx 
f(\infty) \frac{\lambda_{\rm J} }{2r}. \nonumber
\end{eqnarray}
Note that:
\begin{eqnarray}
f(\infty) = \int_0^\infty J_0(x) dx = 1.
\end{eqnarray}
Thus, for $r \ll \lambda_{\rm J} $ we have:
\begin{eqnarray}
{\cal K}_{\rm J} \approx \frac{1}{4\pi \lambda_{ab} \lambda_c r}.
\end{eqnarray}

\end{document}